\documentclass{article}

\usepackage{arxiv}

\usepackage[utf8]{inputenc} 
\usepackage[T1]{fontenc}    
\usepackage{hyperref}       
\usepackage{url}            
\usepackage{booktabs}       
\usepackage{amsfonts}       
\usepackage{nicefrac}       
\usepackage{microtype}      
\usepackage{lipsum}
\usepackage{graphicx}
\graphicspath{ {./images/} }

\usepackage{graphicx}
\usepackage{amsmath}
\usepackage{appendix}
\usepackage{hyperref}
\usepackage{hyperref}
\hypersetup{
    colorlinks=true,
    linkcolor=blue,
	citecolor=green,
}

\usepackage{amsmath,amssymb}
\usepackage[ruled,vlined,lined,commentsnumbered]{algorithm2e}
\usepackage{xspace} 
\usepackage[square,numbers]{natbib}
\usepackage{hyperref}
\hypersetup{colorlinks,allcolors=black}

\title{Analysis of COVID-19 evolution based on testing closeness of sequential data}

\author{
 Tomoko Matsui \\
  The Institute of Statistical Mathematics \\
  \texttt{tmatsui@ism.ac.jp} \\
   \And
 Nourddine Azzaoui  \\
  University of Clermont Auvergne\\
  \texttt{nourddine.azzaoui@uca.fr} \\
  \And
 Daisuke Murakami  \\
  The Institute of Statistical Mathematics \\
  \texttt{dmuraka@ism.ac.jp} \\
}

\begin{document}
\maketitle
\begin{abstract}
A practical algorithm has been developed for closeness analysis of sequential data that combines closeness testing with algorithms based on the Markov chain tester. It was applied to reported sequential data for COVID-19 to analyze the evolution of COVID-19 during a certain time period (week, month, etc.). 
\end{abstract}


\section{Introduction}

The COVID-19 coronavirus has spread worldwide, and as of May 31, 2021, the number of confirmed cases was 170M, and the number of deaths was 3.54M. A fourth wave of infections due to the emergence of variants with strong infectivity began hitting a number of countries in Spring 2021. Coping with a worldwide pandemic like the COVID-19 one requires understanding the infection situation. This requires development of techniques for analyzing the various types of sequential data that are available. These data include the number of confirmed infections, the number of deaths, and the number of polymerase chain reaction tests and rapid antigen tests by location and time.

As the availability of various types of data has increased in recent years, faster and more sample-efficient algorithms have been developed for statistical testing. In particular, for data collected by sensors, closeness testing of distributions to infer information from the underlying probability distributions is rapidly evolving\cite{chanOptimalAlgorithmsTesting2014a, canonneSurveyDistributionTesting2015, daskalakisWhichDistributionDistances2017}. Wolfer and Kontorovich, for example, developed an identity tester that determines whether sequential data represented by two Markov chains are identical\cite{wolferMinimaxTestingIdentity2020}. Although the theory is quite rich in this area, there have been few reports of proposed algorithms being tested on actual applications or of simulation studies. Moreover, the algorithms are suitable only for discrete distributions, so a quantization technique is needed to transform continuous distributions into discrete ones. Canonne and Wimmer  discussed the difficulties inherent in binning and segmentation and their limitations\cite{canonneTestingDataBinnings2020a}.   

We have developed a practical algorithm for closeness analysis of sequential data by combining distribution testing and algorithms based on Wolfer and Kontorovich’s identity tester \cite{wolferMinimaxTestingIdentity2020}. We tested it by using it to analyze the evolution of COVID-19 during a certain time period (week, month, etc.).

In the following section, we briefly describe related work on distribution testing and Markov chain testing. Our analysis methods are described in section \ref{sec:ana}, and their usage for analyzing spatio-temporal data like that for COVID-19 is described in section \ref{sec:exp}. We discuss the testing sensitivity in section \ref{sec:dis} and conclude with a summary of the key points in section \ref{sec:con}.

\section{Related work} \label{sec:rel}

\subsection{Distribution testing} \label{ssec:dist}

Distribution testing typically involves three types of problems: the uniform testing problem, the identity testing problem, and the closeness testing problem. Let $D$ be a distribution over a (countable) domain $\Omega$. The uniform testing problem is to determine whether $D = U_{\Omega}$ (the uniform distribution on $\Omega$) or the distance between $D$ and $U_{\Omega}$ is far from $\varepsilon \in (0,1)$ ($\varepsilon$-far) \cite{batuTestingRandomVariables2001, goldreichTestingExpansionBoundedDegree2011, paninskiCoincidenceBasedTestUniformity2008}. The identity testing problem is to determine whether $D = D^*$ (a fixed distribution over $\Omega$) or $D$ is $\varepsilon$-far from $D^*$ \cite{valiantPowerLinearEstimators2011, valiantAutomaticInequalityProver2014}. The closeness testing problem is to determine whether $D$ and $D'$ (another distribution on $\Omega$) are equal or $\varepsilon$-far from each other \cite{batuTestingClosenessDiscrete2013, valiantTestingSymmetricProperties2011a}. Here, we focus on closeness testing as it is useful for analyzing the the COVID-19 situation. The resulting problem is as follows.

\begin{description}
\item[~~~] Given sample access to distributions $D$ and $D'$ over $\Omega$, and bounds $\eta_1 \geq 0$, $\eta_2 > 0$, $\delta \in (0,1)$, distinguish with probability at least $1 - \delta$ between $d_1(D, D') \leq \eta_1$ and $d_2(D, D') \geq \eta_2$ whenever $D, D'$ satisfy one of these two inequalities.
\end{description}

\noindent Here, $d_1$ and $d_2$ are the distances between two distributions. Depending on the purpose of the analysis, the total variation distance, $l_2$, the $\chi^2$ distance, or the Hellinger distance are generally used as $d_1$ and $d_2$ in distribution testing. The total variation distance is standard, and the properties of the other two distances have been theoretically and comparatively studied \cite{daskalakisWhichDistributionDistances2017}. The $\chi^2$-type statistics defined by Chan et al. \cite{chanOptimalAlgorithmsTesting2014a} are used here.  


\subsection{Markov chain testing} \label{ssec:mc}

Learning and testing discrete distributions has been a hot research area, especially for sample complexity problems in identity testing and closeness testing\cite{canonneSurveyDistributionTesting2015}. Most of the work in this area has relied on independent and identically distributed (iid) sample testing, which is based on an unrealistic assumption. Emergent work has started to address the three testing problems described above, especially for data generated from a finite Markov chain (e.g., \cite{wolferEstimatingMixingTime2019, wolferMinimaxTestingIdentity2020}). Since COVID-19 data observations are obviously not iid in time and space, we assume here that the observed proportions $\pi$ (where the distribution $D$ is estimated by $\pi$) are generated by  a Markov chain over a discrete state space $[s]=\left\{s_1, \dots, s_B\right\}$; this means that it verifies the Markovian property
\begin{equation}\label{MC}
\mathbb{P}\left(\pi_{t}=s_j \mid \pi_{t-1}=s_i\right)=p_{i j} , \quad \hbox{ for all } t,
\end{equation}
where $p_{i j}$ denotes the transition probability from state $s_i$ to state $s_j$. Given an observed trajectory $\boldsymbol{\pi}=\left(\pi_{0}, \ldots, \pi_{T}\right)$ from some unknown Markov chain up to time $T$, we are interested in learning the transition probabilities from only this trajectory. Two strategies can be adopted for Markov chain testing: (i) naive use of distribution testing techniques (closeness testing, identity testing, and so on) for conditional transition probability comparison and (ii) less obvious comparison of the stationary distributions of the two Markov chains. With the first strategy, the discrete conditional probability distributions $ p_{i .}=(p_{i 1}, \ldots, p_{i B})$ and $q_{i .}=(q_{i 1}, \ldots, q_{i B})$ as defined in (\ref{MC}) are compared for each fixed state $s_i$. With the second strategy, this technique needs existence conditions through mixing time concept.

Wolfer and Kontorovich’s identity tester \cite{wolferMinimaxTestingIdentity2020} constructs a tester $\mathcal{T}$ that can determine whether a given trajectory was generated from an unknown ergodic Markov chain $M$ having $B$ states. They showed that the tester can determine with a probability of at least $1-\delta$ whether the sample trajectory was generated from $M$ or $\varepsilon$-far from $M$.

\section{Analysis methods using distribution testing and Markov chain testing} \label{sec:ana}


Focusing on COVID-19, we investigated whether the pandemic evolved in the same way in different regions and for different segments of the population. We tested three analysis methods based on distribution testing and Markov chain testing that can be applied to the spatio-temporal data of COVID-19 and potentially any novel coronavirus. 
\begin{enumerate}
 \item Closeness analysis
 \item Periodical evolution analysis
 \item Key factor analysis
\end{enumerate}
In the following sections, we first formulate the problem and then describe these analysis methods.

\subsection{Observation model formulation} \label{sec:obs}
Let us consider a population $\mathcal{P}$ and suppose that $\mathcal{P} = \bigcup P_l$, where $\{P_l\}_{l=1,\dots, L}$ is a certain segment of the population. This segment can be linked to geographic regions, socio-demographics categories, age, and other relevant auxiliary variables. We are interested in monitoring the dynamic distribution of a coronavirus like COVID-19. We are especially interested in the evolution of the distribution $D_l(t)$ of the number of infected people in population $P_l$ at time $t$.

Our testing framework is applicable to only discrete distributions, so we need to quantize the state space into $B$ bins. Let us denote the discretized states as $[s]=\left\{s_1, \dots, s_B\right\}$ (in the univariate case), and discretization of the interval $[0,p_{max}]$, where $p_{max}$ is the maximum allowed proportion (in the experiments, the segmentation is uniform and $p_{max}$ is less than 1). To investigate the severity of COVID-19, the proportion $\pi_t^l$ of infected people in segment $P_l$ at time $t$ is assigned a state $s_i$ if $ s_i< \pi_t^l \leq s_{i+1} $. The observed proportion is $\hat{\pi}_t^l = {n_t^l}/{N_l}$, where $n_t^l$ is the number of infected people in population $P_l$ at time $t$, and $N_l$ is the size of the population segment $P_l$. For each $t$ and $l$, the application $\hat{\pi}^l_t : \longrightarrow \mathcal{M}[s]$ is to take a random variable in $\mathcal{M}[s]$, which is the set of discrete probability measures on $[s]$.

\subsection{Closeness analysis}

We designed an algorithm for closeness analysis by combining distribution testing (closeness testing) and Markov chain testing in order to analyze the closeness of two sequential data. In distribution testing, there is generally assumed to be oracle access to the distributions. For closeness testing, according to Theorem 1 of Chan et al. \cite{chanOptimalAlgorithmsTesting2014a} and Theorem 3.2.9 of Canonne \cite{canonneSurveyDistributionTesting2015}, tight upper ${\rm O}$ and lower $\Omega$ bounds for sample complexity are given by
$${\rm O}(max(\dfrac{B^{2/3}}{\varepsilon^{4/3}}, \dfrac{B^{1/2}}{\varepsilon^{2}})) \text{ and } \Omega(max(\dfrac{B^{2/3}}{\varepsilon^{4/3}}, \dfrac{B^{1/2}}{\varepsilon^{2}})).$$
The algorithm we designed for closeness analysis satisfies the following two conditions under the assumption of oracle access \cite{canonneSurveyDistributionTesting2015, chanOptimalAlgorithmsTesting2014a}.
On input $\varepsilon \in (0,1)$ (a constant), $C \in \mathbb{R}^+$ (an absolute constant) and $B \in \mathbb{N}$ (the number of states), it takes $C \cdot max(\dfrac{B^{2/3}}{\varepsilon^{4/3}}, \dfrac{B^{1/2}}{\varepsilon^{2}})$ samples from the distributions and, 

\begin{itemize}
 \item if the distributions are equal, it outputs ACCEPT with probability at least ${2}/{3}$;
 \item if the total variation distance between the distributions is greater than $\varepsilon$, it outputs REJECT with probability at least ${2}/{3}$.
\end{itemize}

As shown in Algorithm \ref{alg:test}, five parameters are input: $\varepsilon$, $C$, $B$, $N \in \mathbb{N}$ (the number of testing iterations) and $\mu \in \mathbb{N}$ (the minimum number of samples for testing). The sequential data ($\bf{x}$ and $\bf{y}$ with $d$-dimension) are first quantized into $B$ bins (or $B$ states). Algorithm \ref{alg:test} follows the naive use strategy described in section \ref{ssec:mc}. For each state $b$, the discrete conditional probability distributions ($ p_{b .}=(p_{b 1}, \ldots, p_{b B})=(\frac{T^{x}(1)}{\sum_{k=1}^{B} T^{x}(k)}, \ldots, \frac{T^{x}(B)}{\sum_{k=1}^{B} T^{x}(k)})$ and $q_{b .}=(q_{b 1}, \ldots, q_{b B})=(\frac{T^{y}(1)}{\sum_{k=1}^{B} T^{y}(k)}, \ldots, \frac{T^{y}(k)}{\sum_{k=1}^{B} T^{y}(k)})$) are compared. In accordance with Theorem 1 of Chan et al. \cite{chanOptimalAlgorithmsTesting2014a} and Theorem 3.2.9 of Canonne \cite{canonneSurveyDistributionTesting2015}, $m_0$ is sampled from a Poisson distribution with mean $m$ (line \ref{m}), and $m_0$ samples are sampled from the distributions (lines \ref{tx} and \ref{ty}). For the acceptance probability, the $\chi^2$-type statistic $z(n)$ defined by Chan et al. is calculated for each sample $n$ (line \ref{z}) and compared with a threshold \cite{canonneSurveyDistributionTesting2015} (line \ref{thresz}). The statistic can be viewed as a modification of the empirical triangle distance applied to $c^{x}$ and $c^{y}$. For the reject probability, the total variation distance $d(n)$ is calculated for each sample $n$ (line \ref{d}) and compared with a threshold $\varepsilon$. 
 
After application of Algorithm \ref{alg:test}, the acceptance $P_A$ and reject $P_R$ probabilities, the distance of the $\chi^2$-type statistic $Z$, and the total variation distance $D$ for closeness testing between $\bf{x}$ and $\bf{y}$ can be calculated as the mean, median, or minimum value over all states. The minimum value is the most conservative; the mean value was used in the experiments. The $\chi^2$-type statistic is an estimate of $\chi^2$-divergence. The relation between the divergence and the total variation distance is as follows; for distributions $p$ and $q$, the following inequalities hold.
$$
d_{\mathrm{H}}^{2}\left(p, q\right) \leq d_{\mathrm{TV}}\left(p, q\right) \leq \sqrt{2} d_{\mathrm{H}}\left(p, q\right) \leq \sqrt{d_{\chi^{2}}\left(p, q\right)}.
$$
\noindent Additional details and discussion can be found elsewhere (\cite{daskalakisWhichDistributionDistances2017} for instance). These inequalities show that the $\chi^2$-divergence $d_{\chi^{2}}$ is more conservative than the Hellinger distance $d_{\mathrm{H}}$ and the total variation distance $d_{\mathrm{TV}}$. This motivated our use of the $\chi^2$-type statistic.

\begin{algorithm}
\caption{Closeness analysis of sequential data.}
\label{alg:test}
\LinesNumbered
\SetAlgoLined
\SetKwFunction{Quantize}{Quantize}
\SetKwFunction{Set}{Set}
\SetKwFunction{Sample}{Sample}
\SetKwFunction{Count}{Count}
\SetKwFunction{Test}{Test}
\KwIn{$\varepsilon \in (0,1), C \in \mathbb{R}^+, N \in \mathbb{N}, B \in \mathbb{N}, \mu \in \mathbb{N}$}
\KwData{${\bf x} = (x_{1}, x_{2}, \ldots, x_{I}), {\bf y} = (y_{1}, y_{2}, \ldots, y_{J}) \in \mathbb{R}^d$}
\KwOut{acceptance probability $P_{A}$, reject probability $P_{R}$, $\chi^2$-type statistic $Z$, total variation distance $D$ for each state} 
\tcc{\Quantize ${\bf x}$ and ${\bf y}$ into $B^d$ bins (or $B^d$ states of Markov chains)} 
$Q^{x} = (q^{x}_{1}, q^{x}_{2}, \ldots, q^{x}_{I})$ $\leftarrow {\bf x}$\\
$Q^{y} = (q^{y}_{1}, q^{y}_{2}, \ldots, q^{y}_{J})$ $\leftarrow {\bf y}$\\
$P_{A},P_{R}, Z, D  \leftarrow {\bf 0} \in \mathbb{R}^{B^d}$ \\
\tcc{\Test closeness for each state $b$ }
\For{$b \leftarrow 1$ \KwTo $B^d$} {
$T^{x}$, $T^{y}$ $\leftarrow$ ${\bf 0} \in \mathbb{R}^{B^d}$ \\
$Accept, Reject \leftarrow 0 \in \mathbb{R}$ \\
\tcc{\Count transitions from state $b$ } 
   \For{$i \leftarrow 1$ \KwTo $I-1$} {
     $T^{x}(q^{x}_{i+1}) \leftarrow T^{x}(q^{x}_{i+1}) + {\bf 1}\{q^{x}_{i} = b\}$
   }
   \For{$j \leftarrow 1$ \KwTo $J-1$} {
     $T^{y}(q^{y}_{j+1}) \leftarrow T^{y}(q^{y}_{j+1}) + {\bf 1}\{q^{y}_{j} = b\}$
   }
\tcc{\Test $N$ times}
\If {$(\|T^{x}\|_1 > \mu)$ $\&$ $(\|T^{y}\|_1 > \mu)$}{
\For{$n \leftarrow 1$ \KwTo $N$}{
\Set a variable $m \leftarrow C \cdot max(\frac{B^{2/3}}{\varepsilon^{4/3}}, \frac{B^{1/2}}{\varepsilon^{2}})$  \nllabel{m}\\[4pt]
\Sample variables $m_{0}$ from Poisson distribution with mean $m$ \\[2pt]
\Sample a set $S^{x}$ of $m_{0}$ samples from Markov chain with transition probability $\frac{T^{x}(k)}{\sum_{k=1}^{B} T^{x}(k)}$ \nllabel{tx}\\
\Sample a set $S^{y}$ of $m_{0}$ samples from Markov chain with transition probability $\frac{T^{y}(k)}{\sum_{k=1}^{B} T^{y}(k)}$ \nllabel{ty}\\
\For{$b \leftarrow 1$ \KwTo $B^d$} {
   $c^{x}_{b} \leftarrow \sum_{s \in S^{x}}{{\bf 1}\{s = b\}}$ $, \;$ 
   $c^{y}_{b} \leftarrow \sum_{s \in S^{y}}{{\bf 1}\{s = b\}}$ }
${\it z}(n) \leftarrow \sum^{B^d}_{b=1}{\frac{(c^{x}_{b} - c^{y}_{b})^2 - (c^{x}_{b}+ c^{y}_{b})}{c^{x}_{b} + c^{y}_{b}}}$ $, \;$ \nllabel{z} \\
${\it d}(n) \leftarrow \frac{1}{2}\sum^{B^d}_{b=1}{|\frac{c^{x}_{b}}{\sum^{B^d}_{b=1}c^{x}_{b}} - \frac{c^{y}_{b}}{\sum^{B^d}_{b=1}c^{y}_{b}}|}$ \nllabel{d} \\[4pt]
\If{${\it z}(n) \leq \frac{1}{8}\frac{m^2 \varepsilon^2}{m + B^d}$ \nllabel{thresz}}{\vspace{4pt} $Accept = Accept + 1$} 
\If{${\it d}(n) > \varepsilon$ \nllabel{thresd}}{\vspace{4pt} $Reject = Reject + 1$}
}
$P_{A}(b) \leftarrow \frac{Accept}{N}$ $, \;$ 
$P_{R}(b) \leftarrow \frac{Reject}{N}$ \\[4pt]
$Z(b) \leftarrow \frac{\sum_{n=1}^{N}z(n)}{N}$ $,\;$ 
$D(b) \leftarrow \frac{\sum_{n=1}^{N}d(n)}{N} $
}
\Else
{
$P_{A}(b) = P_{R}(b) = Z(b) =  D(b) = -1$
}
}
\end{algorithm}

\subsection{Periodical evolution analysis}
For a sequential data such as COVID-19 data, it is often demanded to analyze the evolution situation. Here, we investigate a method of periodical evolution analysis with closeness analysis. As shown in Algorithm \ref{alg:evo}, input sequence ${\bf x}$ is first segmented into $L$ segments. Then, for each pair of segments, closeness of the pair is tested using Algorithm \ref{alg:test}. We can analyze the periodical properties on the resulting $L \times L$ matrices for the acceptance probabilities and the distances.
\begin{center}
\begin{algorithm}[H]
\LinesNumbered
\SetAlgoLined
\SetKwFunction{Quantize}{Quantize}
\SetKwFunction{Segment}{Segment}
\SetKwFunction{Set}{Set}
\SetKwFunction{Sample}{Sample}
\SetKwFunction{Count}{Count}
\SetKwFunction{Test}{Test}
\KwIn{$\varepsilon \in (0,1), C \in \mathbb{R}^+, N \in \mathbb{N}, B \in \mathbb{N}$, $L \in \mathbb{N}$}
\KwData{${\bf x} = (x_{1}, x_{2}, \ldots, x_{I}) \in \mathbb{R}^d$}
\KwOut{acceptance probability $P_{A}$, reject probability $P_{R}$, $\chi^2$-type statistic $Z$, total variation distance $D$ for each pair of segments} 
\tcc{\Segment ${\bf x}$ into $L$ segments}
$ {\bf x} = \{ {\bf x}^1, {\bf x}^2, \ldots, {\bf x}^L\}$ \\
\tcc{\Test closeness for each pair of segments }
\For{$i \leftarrow 1$ \KwTo $L$}{
\For{$j \leftarrow 1$ \KwTo $L$}{
\If{$i \neq j$}{
\tcc{\Test closeness of segments ${\bf x}^i$ and ${\bf x}^j$ using Algorithm \ref{alg:test}}
}
}
}
\caption{Periodical evolution analysis using testing closeness.}
\label{alg:evo}
\end{algorithm}
\end{center}

\clearpage

\subsection{Key factor analysis} \label{ssec:key}

When planning measurements such as those for COVID-19, it is important to analyze the key factors, i.e., the factors that correlate with changes in, for example, the number of infections. We investigated a method for analyzing the key factors that uses a generalized additive model (GAM) \cite{gam1990} in which the response variable depends linearly on the unknown smooth functions of some predictor variables and the focus is on making inferences about the smooth functions. The benefit of GAM is that it takes advantage of the smoothed transforms of the predictor variables using basis functions such as smoothing splines. The distances obtained by the closeness analysis are used as the response variables. The data for the key factor candidates, e.g., vehicle and public transport increase rates, are used as predictor variables. The best model is then selected in a step-wise fashion using either Akaike Information Criterion or model residual deviance \cite{gam1992}. 
 
\section{Experiments and results} \label{sec:exp}
\subsection{COVID-19 sequential data}
We used reported data for the number of newly infected people $n_l^t$ for each of the 53 cities on the main island of Japan as reported daily by the Tokyo metropolitan government from April 1, 2020, to May 6, 2021, along with the population $N_l$ of each city. Segmentation $\{P_l\}_{l=1,\dots, L}$ (described in section \ref{sec:obs}) was linked to each city in Tokyo (which is a prefecture, not a city). The observed proportion $\hat{\pi}_t^l (= {n_t^l}/{N_l})$ was quantized into $B$-states, and $B$ was set to 20. 

\begin{figure}
  \includegraphics[width=\linewidth]{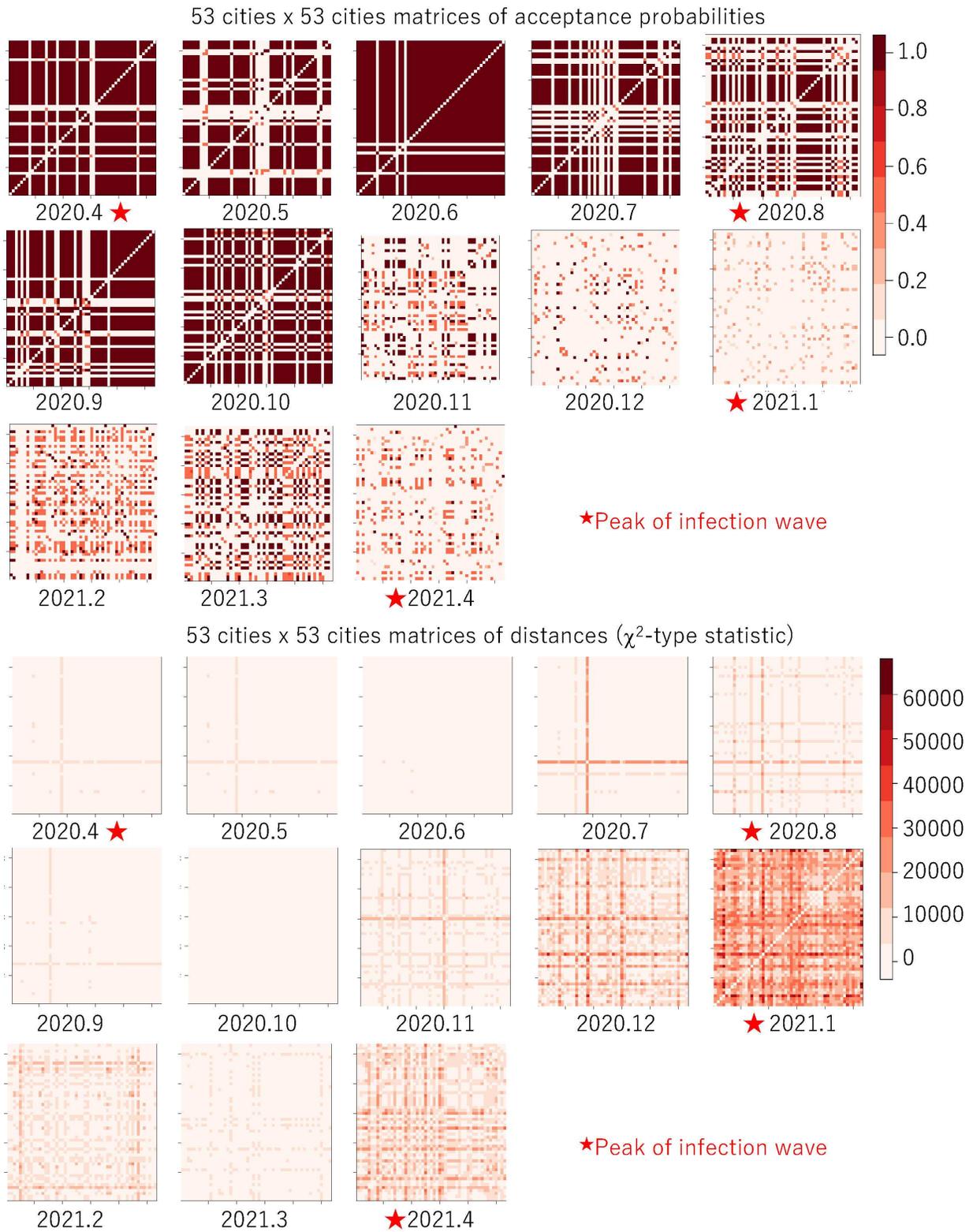}
\caption{Acceptance probabilities (top) and distances (bottom) for closeness analysis of COVID-19 infection status between 53 cities in Tokyo.}
\label{fig:1}       
\end{figure}

\begin{figure}
  \includegraphics[width=\linewidth]{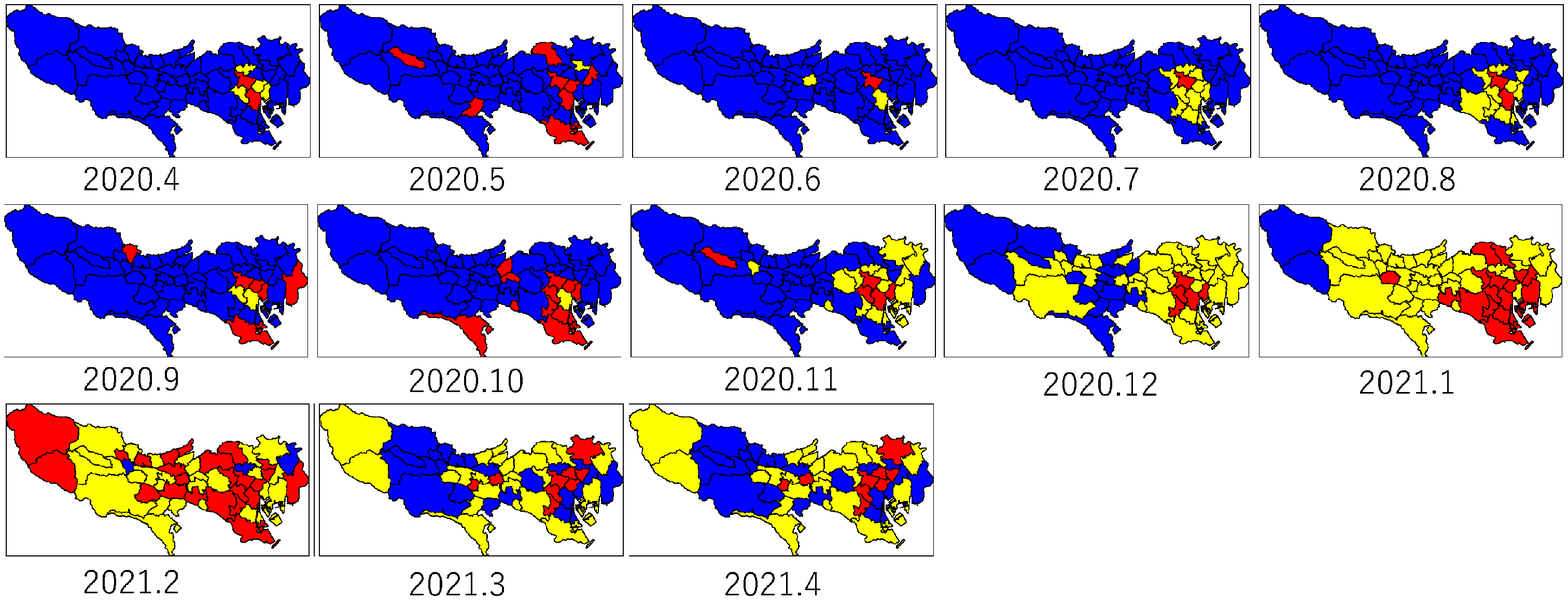}
\caption{k-means clustering for 53 cities in Tokyo by month based on distance matrices: red indicates relatively high level of increases in infection, yellow indicates moderate level, and blue indicates low level.}
\label{fig:2}       
\end{figure}

\subsection{Closeness analysis of COVID-19 infection situation between cities} \label{ssec:close}

Figure \ref{fig:1} shows $53$ cities $\times 53$ cities matrices of acceptance probabilities (the mean of $P_A(b)$ over all states in Algorithm \ref{alg:test}) and distances of $\chi^2$-type statistics (the mean of $Z(b)$ over all states in Algorithm \ref{alg:test}) between all pairs of 53 cities in Tokyo for each month from April 2020 to April 2021, calculated using Algorithm \ref{alg:test}. As of June 2021, there had been four waves of COVID-19 infection; the peak months are roughly indicated by red stars.

For the acceptance probabilities, the matrices between the waves tend to be darker; that is, many cities are considered to have had similar characteristics of the changes in the number of infected people for each of the months. In fact, for such cities, the number of infected people was relatively and stably small during those months. 

For the distances, the overall matrix color is the darkest for January 2021, when the third wave peaked and the number of infected people was the largest. Many cities experienced an explosion of infections and different characteristics of the changes in the number of infected people for the month. 

Figure \ref{fig:2} shows the k-means clustering for the distance matrices in Figure \ref{fig:1}. 
To facilitate recognition of the differences in the level of increases in infection, the number of color codes was set to three: red indicates relatively high level, yellow indicates moderate level, and blue indicates low level.
For April 2020, two cities in the heart of Tokyo, Shinjuku-ku and Minato-ku, had the highest level. This is attributed to Shinjuku-ku and Minato-ku having a popular entertainment district. Until October 2020, most cities had the lowest level. Starting with the third wave, roughly from December 2020 to February 2021, the levels of the nearby cities increased to moderate and then to high. These figures illustrate how the characteristics of the changes in the number of infected people were transformed.

\begin{figure}
  \includegraphics[width=\linewidth]{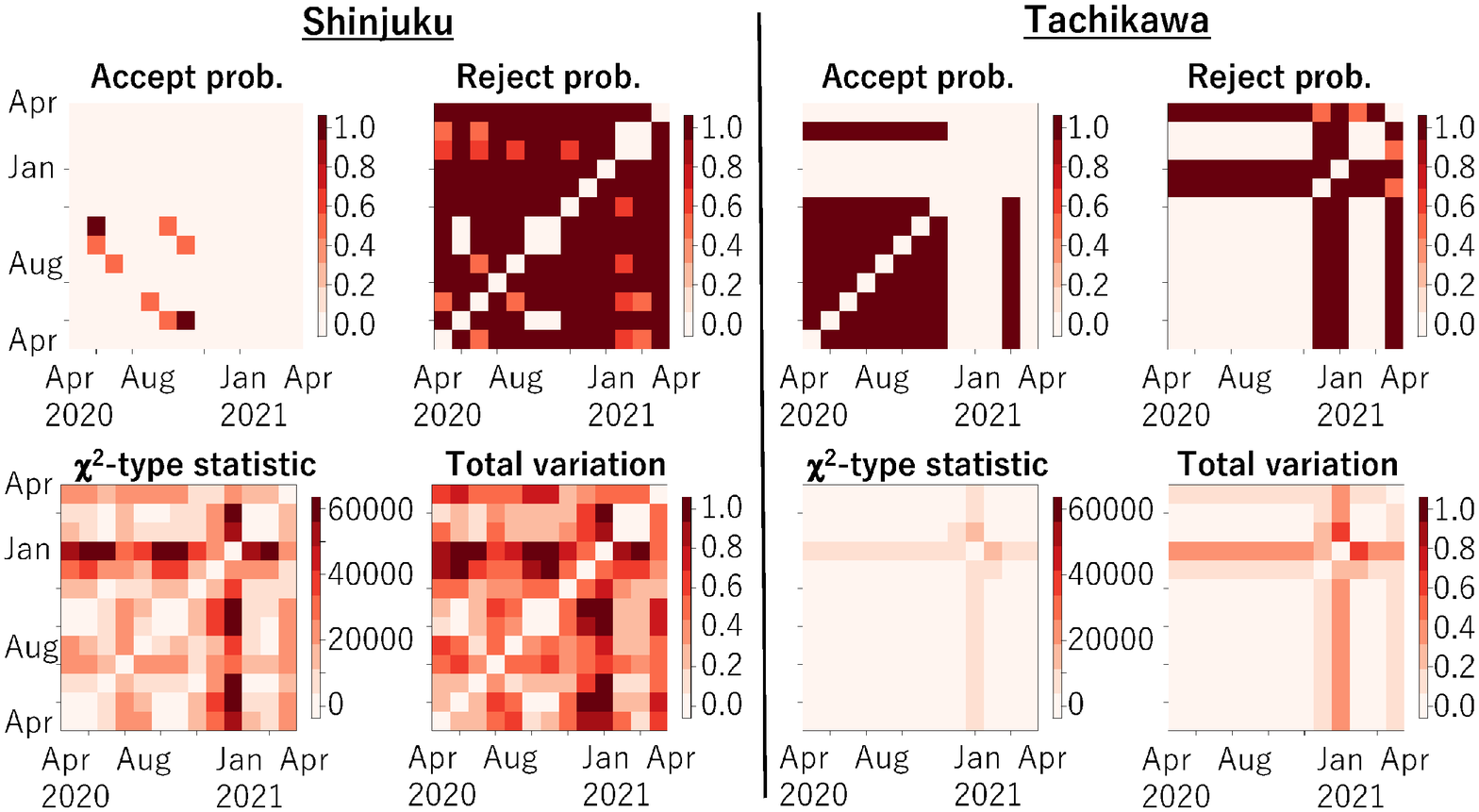}
\caption{For Shinjuku (left) and Tachikawa (right), $13 \text{months}\times 13 \text{months}$ matrices of acceptance probabilities, distance of $\chi^2$-type statistic, reject probability, and total variation distance.}
\label{fig:3}       
\end{figure}

\begin{figure}
  \includegraphics[width=\linewidth]{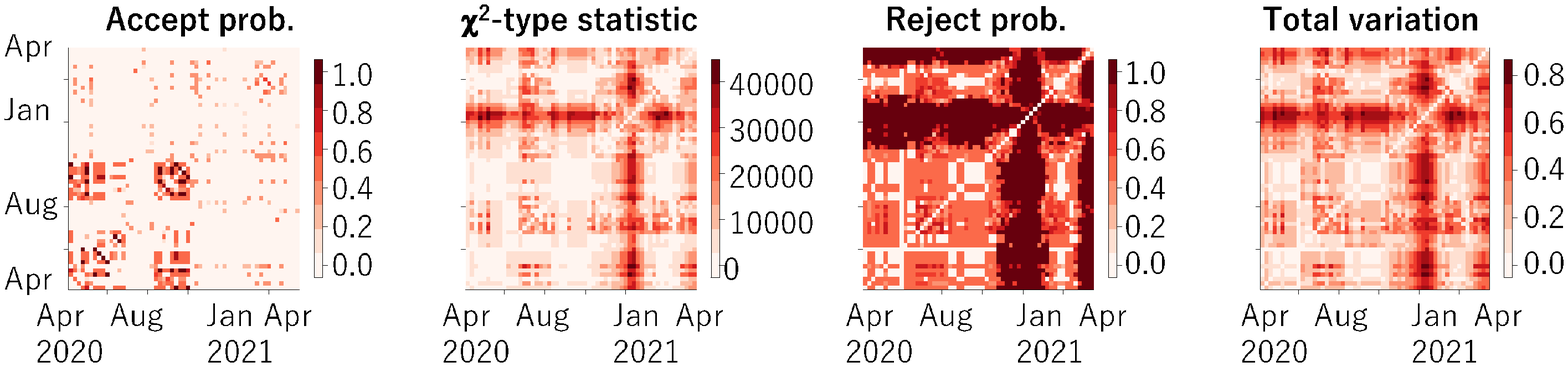}
\caption{For all cities in Tokyo, $57 \text{weeks}\times 57 \text{weeks}$ matrices of acceptance probabilities, distance of $\chi^2$-type statistic, reject probability, and total variation distance.}
\label{fig:4}       
\end{figure}

\subsection{Periodical COVID-19 evolution analysis}

Figure \ref{fig:3} shows the matrices of acceptance probabilities, distances of $\chi^2$-type statistics, reject probabilities (mean of $P_R(b)$ over all states in Algorithm \ref{alg:test}), and total variation distances (mean of $D(b)$ over all states in Algorithm \ref{alg:test}) between all pairs of 13 months for Shinjuku and Tachikawa calculated using Algorithm \ref{alg:evo}. Tachikawa-shi is located in the middle west of Tokyo, in a suburban area. For Shinjuku-ku (in the heart of Tokyo), as in Figure \ref{fig:1}, almost all the pairs are different while the May–October 2020 pair are similar. For Tachikawa-shi, the pairs from April to November 2020 and for February and March 2021 are similar. The number of infected people for these months was relatively and stably small. This figure illustrates the characteristics of monthly COVID-19 evolution for both cities.

Figure \ref{fig:4} shows the matrices of acceptance probabilities, distances of $\chi^2$-type statistic, reject probabilities, and total variation distances between all pairs of 57 weeks from 1 April 2020 to 5 May 2021 for all of Tokyo calculated using Algorithm \ref{alg:evo} and all the numbers accumulated for all the cities in Tokyo. The acceptance probabilities show that the weeks from April to June, 2020 and for August and September, 2020, tended to be similar among the cities. The distances show that the weeks in January, April, and May 2021 were very different. This indicates that the number of infected people for the weeks in January 2021 dynamically changed, probably because of an increase in contacts between people due to year-end and beginning-of-year parties and meetings. In April and May 2021, variants of the COVID-19 virus with higher infectivity began to gradually spread, so the characteristics of the changes in the number of infected people differed from those in previous weeks.

\subsection{Key factor analysis for COVID-19 evolution}

For the key factor analysis, we used the distances of the $\chi^2$-type statistic $Z$ and the total variation distances $D$ between all pairs of 52 weeks from 6 May 2020 to 4 May 2021 for all of Tokyo, which are included in figure \ref{fig:4} in which 57 weeks were used. Table \ref{tab:kf} lists the key factor candidates used in the experiments such as vehicle and public transport increase rates and average temperature in Tokyo, which are considered to affect the rate of new infections. We set a delay of zero (no delay), one week, or two weeks between the distances. 

For the distances of the $\chi^2$-type statistic, the R-squared (adjusted) values are listed in Table \ref{tab:r}. R-squared is a statistical measure of the success in explaining the response by the model, and R-squared (adjusted) is a version adjusted for the number of predictors in the model for parsimony. The table shows that the fitting was fairly accurate. The best model for a delay of two weeks was selected; it is shown in eq. (\ref{eq:z}). The $s(\it term)$ indicates a smoothed transform in which $\it term$ is computed using a smoothing spline, as mentioned in section \ref{ssec:key}. All the terms were significant: $0.001$ significance level for $\bf{vehicle}$, $s(\bf temperature)$, and $s(\bf deathTokyo)$, $0.01$ for $s(\bf week)$, $s(\bf patientHospital)$, and $s(\bf roomHospital)$, and $0.05$ for $\bf{pedestrian}$ and $s(\bf deathWorld)$.

\begin{equation} \label{eq:z}
\begin{split}
   Z \sim & s(\bf week) + \bf{vehicle} + \bf{pedestrian} + s(\bf tepmerature)+ s(\bf deathTokyo) \\
   &  +  s(\bf deathWorld) + s(\bf patientHospital) + s(\bf roomHospital)
\end{split}
\end{equation}

\noindent For the total variation distances, the fitting accuracy on the R-squared (adjusted) values was fairly good, as shown in Table \ref{tab:r}. The best model for a delay of two weeks was selected; it is shown in eq. (\ref{eq:d}). All the terms were significant except for $s(\bf patientHospital)$: $0.001$ significance level for $s(\bf week)$, $\bf vehicle$, $s(\bf temperature)$, $s(\bf deathTokyo)$, and $s(\bf infectedWorld)$ and $0.01$ for $\bf pedestrian$ and $s(\bf room\-Hospital)$.
\begin{equation} \label{eq:d}
\begin{split}
    D \sim &  s(\bf week) + \bf{vehicle} + \bf{pedestrian} + s(\bf temperature) + s(\bf deathTokyo)  \\
    & + s(\bf infectedWorld) + s(\bf patientHospital) + s(\bf roomHospital)
\end{split}
\end{equation}

Moreover, we divided the 52 weeks from 6 May 2020 to 4 May 2021 into two periods: (i) the 30 weeks from May to November 2020 and (ii) the 22 weeks from December 2020 to May 2021. For the first period, the R-squared (adjusted) values for both the $\chi^2$-type statistic and total variation distance in Table \ref{tab:r} were low, making it is difficult to find correlation between the distances and the key factors. For the second period, the R-squared (adjusted) values for both distances were high. As mentioned in section \ref{ssec:close}, the third wave roughly started in December 2020 in Tokyo, and stronger correlations between the distances and the key factors are evident for the second period. 

For the distances of the $\chi^2$-type statistic, the best model for a delay of two weeks was selected; it is shown in eq. (\ref{eq:zlat}). All the terms were significant: $0.001$ significance level for $\bf week$, $\bf{vehicle}$, $s(\bf deathTokyo)$, $s(\bf deathWorld)$, and $s(\bf patientHospital)$, $0.01$ for $s(\bf transport)$ and $\bf infectedWorld$, and $0.05$ for $s(\bf{temperature})$.

\begin{equation} \label{eq:zlat}
\begin{split}
    Z \sim & \bf{week} + \bf{vehicle} + s(\bf transport) + s(\bf temperature) + s(\bf deathTokyo) \\
   &  + \bf{infectedWorld} + s(\bf deathWorld) + s(\bf patientHospital)
\end{split}
\end{equation}

\noindent For the total variation distances, the best model for a delay of two weeks was selected; it is shown in eq. (\ref{eq:dlat}). All the terms were significant except for $s(\bf patientHospital))$: $0.001$ significance level for $\bf week$, $\bf vehicle$, $s(\bf deathTokyo)$, and $s(\bf patientHospital)$ and $0.05$ for $s(\bf transport)$.
\begin{equation} \label{eq:dlat}
\begin{split}
    D \sim &  \bf{week} + \bf{vehicle} +  s(\bf transport) + s(\bf deathTokyo) \\
    & + s(\bf patientHospital)
\end{split}
\end{equation}

These results indicate that the increase rates for vehicles and public transport can be used in the COVID-19 measurements, especially for the second period. The temperature, numbers of deaths, and number of patients in hospitals in Tokyo should be considered key factors that can be correlated with a change in COVID-19 infection rates.

\begin{table}
\begin{center}
\caption{Key factor candidates as predictor variables.}
\label{tab:kf} 
\begin{tabular}{|l|p{8 cm}|}
\hline\noalign{\smallskip}
Predictor variable & Description \\ 
\noalign{\smallskip}\hline\hline\noalign{\smallskip}
\bf week & Time point (weekly ID) \\ 
\noalign{\smallskip}\hline\noalign{\smallskip}
\bf vehicle & Vehicle increase rate (provided by Apple Inc.; compared with January 13, 2020)\\ 
\noalign{\smallskip}\hline\noalign{\smallskip}
\bf transport & Public transport increase rate (provided by Apple Inc.; compared with January 13, 2020)\\ 
\noalign{\smallskip}\hline\noalign{\smallskip}
\bf pedestrian & Pedestrian increase rate (provided by Apple Inc.; compared with January 13, 2020)\\ 
\noalign{\smallskip}\hline\noalign{\smallskip}
\bf temperature & Average temperature in Tokyo (provided by Japan Meteorological Agency)\\ 
\noalign{\smallskip}\hline\noalign{\smallskip}
\bf deathTokyo & Number of COVID-19 deaths in Tokyo (provided by Ministry of Health, Labour and Welfare)\\ 
\noalign{\smallskip}\hline\noalign{\smallskip}
\bf patientHospital & Number of patients in hospitals in Tokyo (provided by Ministry of Health, Labour and Welfare )\\ 
\noalign{\smallskip}\hline\noalign{\smallskip}
\bf roomHospital & Number of available rooms in hospitals in Tokyo (provided by Ministry of Health, Labour and Welfare )\\ 
\noalign{\smallskip}\hline\noalign{\smallskip}
\bf infectedWorld & Number of people infected with COVID-19 worldwide (obtained from Our World in Data)\\
\noalign{\smallskip}\hline\noalign{\smallskip}
\bf deathWorld & Number of COVID-19 deaths in the world (obtained from Our World in Data )\\
\noalign{\smallskip}\hline
\end{tabular}
\end{center}
\end{table}
\begin{table}
\begin{center}
\caption{R-squared (adjusted) values for response variables of distances of $\chi^2$-type statistic and total variation with a delay of zero, one week, or two weeks from time points of predictor variables.}
\label{tab:r} 
\begin{tabular}{|l|r|r|r|r|r|r|}
\hline\noalign{\smallskip}
 Period & \multicolumn{3}{|c|}{$\chi^2$-type statistic} & \multicolumn{3}{c|}{Total variation} \\ 
\noalign{\smallskip}\cline{2-7}\noalign{\smallskip}
 & no delay & 1 week& 2 weeks & no delay & 1 week& 2 weeks\\
\noalign{\smallskip}\hline\hline\noalign{\smallskip}
All 52 weeks & 0.52 & 0.53 & \bf{0.55} & 0.52 & 0.55 & \bf{0.58} \\
\noalign{\smallskip}\hline\noalign{\smallskip}
(i) First 30 weeks & 0.32 & 0.27 & 0.28 & 0.32 & 0.31 & 0.35 \\
\noalign{\smallskip}\hline\noalign{\smallskip}
(ii) Last 22 weeks & 0.51 & 0.60 & \bf{0.70} & 0.63 & 0.59 & \bf{0.67} \\
\noalign{\smallskip}\hline
\end{tabular}
\end{center}
\end{table}


\begin{figure}
  \includegraphics[width=\linewidth]{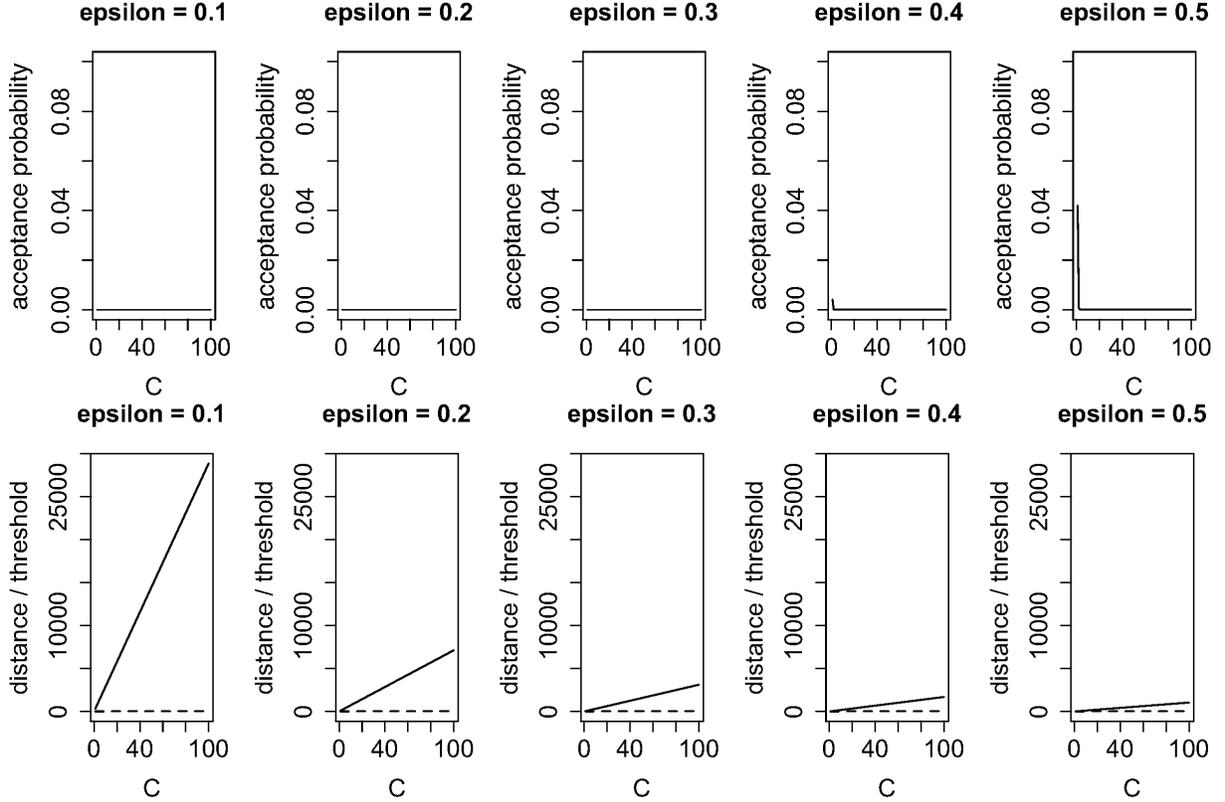}
\caption{Acceptance probabilities and distance of $\chi^2$-type statistic (solid line) and threshold (dashed line) of closeness analysis between two different sequences ($Q^x$ and $Q^y$) with/without Markovian property and with various values of $\varepsilon$ and $C$ in Algorithm \ref{alg:test}.}
\label{fig:5}       
\end{figure}

\begin{figure}
  \includegraphics[width=\linewidth]{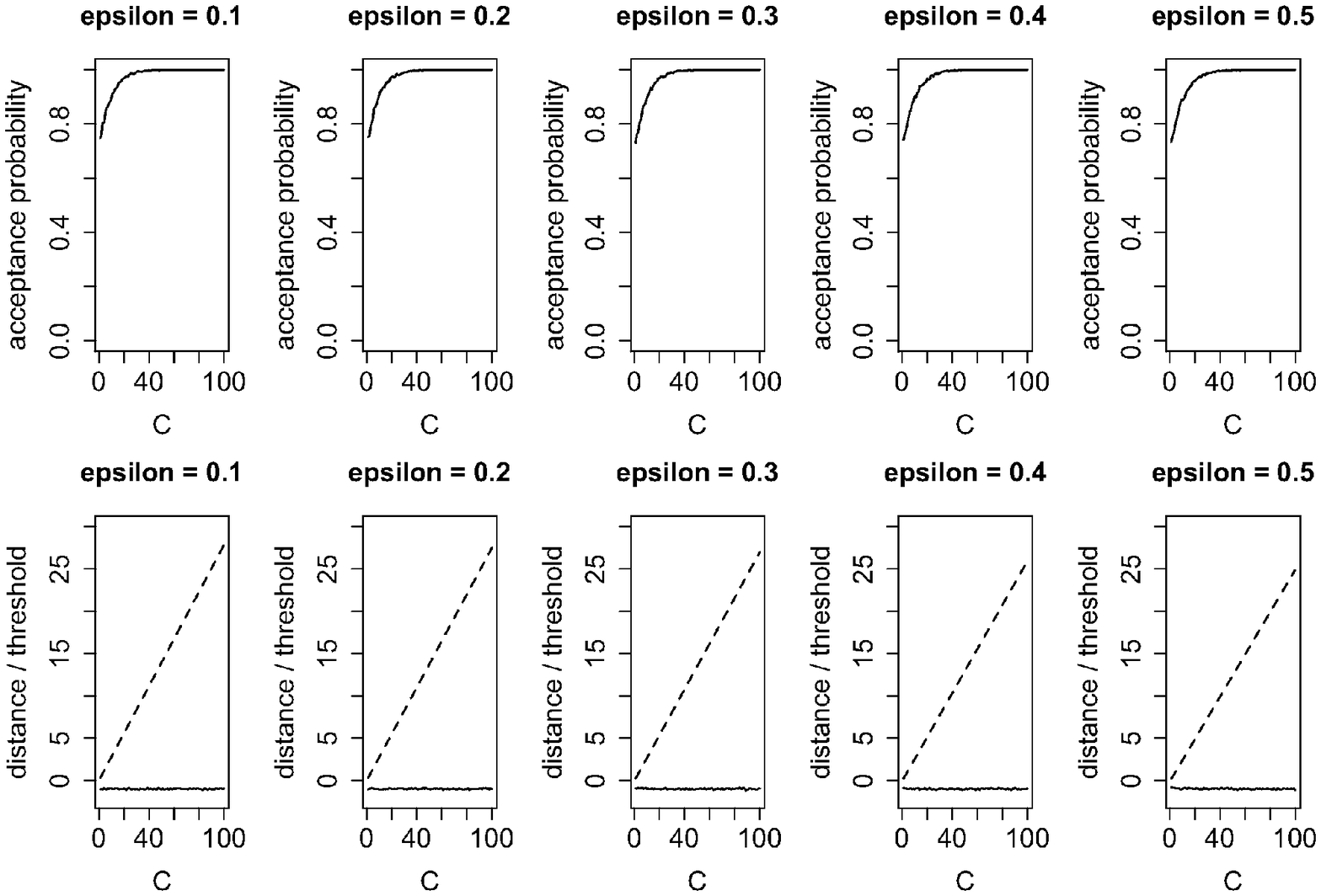}
\caption{Acceptance probabilities and distance of $\chi^2$-type statistic (solid line) and threshold (dashed line) of closeness analysis between two identical sequences ($Q^x$ and $Q^x$) with various values of $\varepsilon$ and $C$ in Algorithm \ref{alg:test}.}
\label{fig:6}       
\end{figure}

\section{Discussion} \label{sec:dis}

We first discuss the properties of Algorithm \ref{alg:test} as a Markov chain tester and the sensitivity of its parameters. We do this using simulated data: (i) sequence $Q^x$ randomly generated from a transition probability matrix with $5$ states (Markov chain), (ii) sequence $Q^y$ generated using sorting sequence $X$, and (iii) sequence $Q^z$ consisting of $(100 - \alpha)$\% sequences (the same as for $Q^x$) and an $\alpha$\% sequence (different from $Q^x$). All sequences had a length of 100 with state components $s_1=1,\ldots,s_5=5$ (see appendix \ref{app:data}). Note that although sequences $Q^x$ and $Q^y$ included the same portion of each state, $Q^y$ had no Markovian property. 

Figure \ref{fig:5} shows the acceptance probabilities, the distances of the  $\chi^2$-type statistic, and the threshold values of closeness analysis between two sequences ($Q^x$ and $Q^y$) with and without the Markovian property and with various values of $\varepsilon$ and $C$ in Algorithm \ref{alg:test}. When $\varepsilon$ was smaller than 0.3, the algorithm could accurately distinguish $Q^x$ and $Q^y$ for all values of $C$. However, when $\varepsilon$ was $0.4$ or $0.5$ and $C$ was 1 or less, the test results were incorrect although the inaccuracy was less than $4\%$. These results show that strict testing can be conducted with small values of $\varepsilon$ and large values of $C$ although with these setting, $m$ (line \ref{m} in Algorithm \ref{alg:test}) becomes large and the computation cost is higher. However, the required level of strictness in closeness analysis should differ between applications, meaning that the values can be set accordingly, especially that of $\varepsilon$. Moreover, both $C$ and $\varepsilon$ should be set in accordance with the available computation power.

Figure \ref{fig:6} shows the acceptance probabilities, the distances of the $\chi^2$-type statistic, and the threshold values of closeness analysis between two identical sequences ($Q^x$ and $Q^x$) with various values of $\varepsilon$ and $C$ in Algorithm \ref{alg:test}. For $\varepsilon$ from 0.1 to 0.9 and $C$ from 1 to 100, the algorithm  correctly determined that the two sequences were the same.

\begin{table}
\begin{center}
\caption{Acceptance probability, distance of $\chi^2$-type statistic, reject probability, and total variation distance of closeness analysis between (100 - $\alpha$)\% similar sequences ($Q^x$ and $Q^z$) with $\varepsilon = 0.1$ and $C = 100$ in Algorithm \ref{alg:test}.}
\label{tab:1} 
\begin{tabular}{|l|r|r|r|r|r|r|}
\hline\noalign{\smallskip}
$\alpha$ & $0\%$ & $1\%$ & $2\%$ & $3\%$ & $4\%$ & $5\%$ \\ 
\noalign{\smallskip}\hline\hline\noalign{\smallskip}
Accept probability &1.0 &0.8 &0.4 & 0.2&0.2 &0.0 \\ 
\noalign{\smallskip}\hline\noalign{\smallskip}
$\chi^2$-type statistic &-1.0 &64.3 &173.5 & 244.0& 301.5& 492.9\\ 
\noalign{\smallskip}\hline\noalign{\smallskip}
Reject probability & 0.0& 0.0&0.0 &0.2 & 0.2& 0.6\\ 
\noalign{\smallskip}\hline\noalign{\smallskip}
Total variation distance & 0.0 & 0.0& 0.0& 0.1& 0.1& 0.1 \\ 
\noalign{\smallskip}\hline\hline\noalign{\smallskip}
Wilcoxon rank-sum test: p-value & 1 & 0.9 & 0.9 & 0.9 & 0.8 & 0.8 \\
\noalign{\smallskip}\hline\noalign{\smallskip}
Kolmogorov-Smirnov test: p-value &1 &1 &1 &1 & 1 & 1 \\
\noalign{\smallskip}\hline
\end{tabular}
\end{center}
\end{table}

Table \ref{tab:1} lists the acceptance probabilities, the distances of the $\chi^2$-type statistic, the reject probabilities, and the total variation distances of closeness analysis between (100 - $\alpha$)\% similar sequences ($Q^x$ and $Q^z$) with $\varepsilon = 0.1$ and $C = 100$ in Algorithm \ref{alg:test}. $\alpha$ was varied from $0$ to $5\%$. The algorithm was able to distinguish the similar sequences when $\alpha = 2\%$ or more. In contrast, the classical hypothesis tests for two distributions (Wilcoxon rank-sum test and Kolmogorov-Smirnov test) could not reject the null hypothesis for all values of $\alpha$. The proposed algorithm thus has strong testing power for sequential data.

\section{Conclusions} \label{sec:con}


We have designed a practical algorithm for testing the closeness of sequential data by combining distribution testing and Markov chain testing. We used it to analyze the closeness, the periodical evolution, and the key factors for the number of people infected with COVID-19 for each city in Tokyo. The results showed that whether or not the epidemic evolves in the same way in different cities or in different months or weeks with numerical indicators of the acceptance and reject probabilities and the significance levels. Examination of the properties of the algorithm as a Markov chain tester and the sensitivity of the parameters showed that strict testing can be conducted with small values of $\varepsilon$ and large values of $C$ under the constraint of the available computation power. Comparison with the classical Wilcoxon rank-sum test and Kolmogorov-Smirnov test demonstrated that the algorithm has a strong testing power for sequential data.



\bibliography{ourrefs}

\section*{Appendix}
\section{Simulated data} \label{app:data}
The simulated data, $Q^x$,  $Q^y$  and $Q^z$  are as follows.
\begin{description}
\item[$\,\,$] $Q^x$ = (1 4 1 2 2 5 1 2 2 5 5 5 1 2 5 5 3 3 4 5 4 2 4 4 5 3 4 4 5 5 5 5 4 3 2 2 5
1 4 3 2 4 5 3 5 5 1 5 2 3 5 3 2 4 1 2 4 4 5 5 1 2 2 1 2 2 1 5 5 3 5 3 5 1
2 4 5 3 4 4 4 5 4 3 1 4 5 4 5 4 3 2 1 3 2 3 5 1 3 4)
\item[$\,\,$] $Q^y$ = (1 1 1 1 1 1 1 1 1 1 1 1 1 1 2 2 2 2 2 2 2 2 2 2 2 2 2 2 2 2 2 2 2 3 3 3 3
3 3 3 3 3 3 3 3 3 3 3 3 4 4 4 4 4 4 4 4 4 4 4 4 4 4 4 4 4 4 4 4 4 4 5 5 5
5 5 5 5 5 5 5 5 5 5 5 5 5 5 5 5 5 5 5 5 5 5 5 5 5 5)
\item[$\,\,$] $Q^z$ = (1 4 1 2 2 5 1 2 2 5 5 5 1 2 5 5 3 3 4 5 4 2 4 4 5 3 4 4 5 5 5 5 4 3 2 2 5
1 4 3 2 4 5 3 5 5 1 5 2 3 5 3 2 4 1 2 4 4 5 5 1 2 2 1 2 2 1 5 5 3 5 3 5 1
2 4 5 3 4 4 4 5 4 3 1 4 5 4 5 4 3 2 1 3 2 {\bf 2 2 2 2 2}) $\,\,\,\,$ ($\alpha = 5\%$)
\end{description}

\noindent The transition probability matrix used to generate $Q^x$ is as follows.

\begin{center}
$$
\begin{pmatrix}
0.02126912 & 0.40209113 & 0.3423650 & 0.1571781 & 0.07709659 \\
0.19377434 & 0.19871080 & 0.1079850 & 0.1904423 & 0.30908763 \\
0.16414480 & 0.33028736 & 0.0176185 & 0.3189076 & 0.16904172 \\
0.04017933 & 0.03392901 &  0.2268634 & 0.2755908 & 0.42343754 \\
0.24338862 & 0.09483701 & 0.2326078 & 0.1308475 & 0.29831911 \\
\end{pmatrix}
$$
\end{center}

\end{document}